\documentclass[twocolumn,showpacs,superscriptaddress,longbibliography,prl]{revtex4-1}

\usepackage{graphicx}
\usepackage{dcolumn}
\usepackage{amsmath}
\usepackage{color}
\usepackage{braket}
\usepackage{float}
\usepackage[english]{babel}

\newcommand{\SrCoVO}{SrCo$_2$V$_2$O$_8$}
\newcommand{\BaCoVO}{BaCo$_2$V$_2$O$_8$}

\begin{document}

\title{Quantum Critical Dynamics of a Heisenberg-Ising Chain in a Longitudinal Field: Many-Body Strings versus Fractional Excitations}

\author{Zhe~Wang}
\affiliation{Institute of Radiation Physics, Helmholtz-Zentrum Dresden-Rossendorf, 01328 Dresden, Germany}

\author{M.~Schmidt}
\author{A.~Loidl}
\affiliation{Experimental Physics V, Center for Electronic
Correlations and Magnetism, Institute of Physics, University of Augsburg, 86135 Augsburg, Germany}

\author{Jianda~Wu}
\author{Haiyuan Zou}
\affiliation{Tsung-Dao Lee Institute \& School of Physics and Astronomy, Shanghai Jiao Tong University, Shanghai 200240, China}

\author{Wang Yang}
\affiliation{Stewart Blusson Quantum Matter Institute, University of British Columbia, Vancouver, Canada}

\author{Chao Dong}
\affiliation{Institute for Solid State Physics, University of Tokyo, Kashiwa, Chiba 277-8581, Japan}
\affiliation{Wuhan National High Magnetic Field Center and School of Physics,
Huazhong University of Science and Technology, Wuhan 430074, China}

\author{Y. Kohama}
\author{K. Kindo}
\affiliation{Institute for Solid State Physics, University of Tokyo, Kashiwa, Chiba 277-8581, Japan}

\author{D.~I.~Gorbunov}
\affiliation{Dresden High Magnetic Field Laboratory (HLD-EMFL), Helmholtz-Zentrum
Dresden-Rossendorf, 01328 Dresden, Germany}

\author{S. Niesen}
\author{O. Breunig}
\author{J. Engelmayer}
\author{T. Lorenz}
\affiliation{Institute of Physics II, University of Cologne, 50937 Cologne, Germany}

\date{\today}

\begin{abstract}
We report a high-resolution terahertz spectroscopic study of quantum spin dynamics in the antiferromagnetic Heisenberg-Ising spin-chain compound BaCo$_2$V$_2$O$_8$ as a function of temperature and longitudinal magnetic field. Confined spinon excitations are observed in an antiferromagnetic phase below $T_N\simeq 5.5~K$. In a field-induced gapless phase above $B_c=3.8$~T, we identify many-body string excitations as well as low-energy fractional psinon/antipsinon excitations by comparing to Bethe-Ansatz calculations. In the vicinity of $B_c$, the high-energy string excitations are found to be dynamically dominant over the fractional excitations.
\end{abstract}

\maketitle

Understanding quantum critical dynamics in strongly correlated many-body systems
is a compelling and challenging task in modern condensed matter physics \cite{Sachdev08,Lnsen07,Gegenwart08,Keimer15,Balents10,Giamarchi08}.
For example, the onset of unconventional superconductivity is proximate to a quantum critical point, which is featured by enhanced spin fluctuations due to suppression of long-range antiferromagnetic order by external parameters, e.g. hole doping in copper oxides \cite{Keimer15} or applied pressure in heavy fermion systems \cite{Gegenwart08}.
In frustrated and/or low-dimensional quantum magnets, even without the complication of charge correlations, exotic spin states, such as quantum spin liquids \cite{Balents10} or a magnonic Bose-Einstein condensate \cite{Giamarchi08}, emerge at quantum phase transitions  \cite{Sachdev08}.
However, the corresponding critical dynamics is generally elusive,
and in particular, the role of high-energy excitations, such as many-body bound states of magnons \cite{Bethe31,Gaudin71,Takahashi72,Takahashi05}, remains to be clarified,
as compared to the well-established low-energy fractionalized spin excitations \cite{Faddeev81,Shiba80}.

\begin{figure}[t]
\centering
\includegraphics[width=80mm,clip]{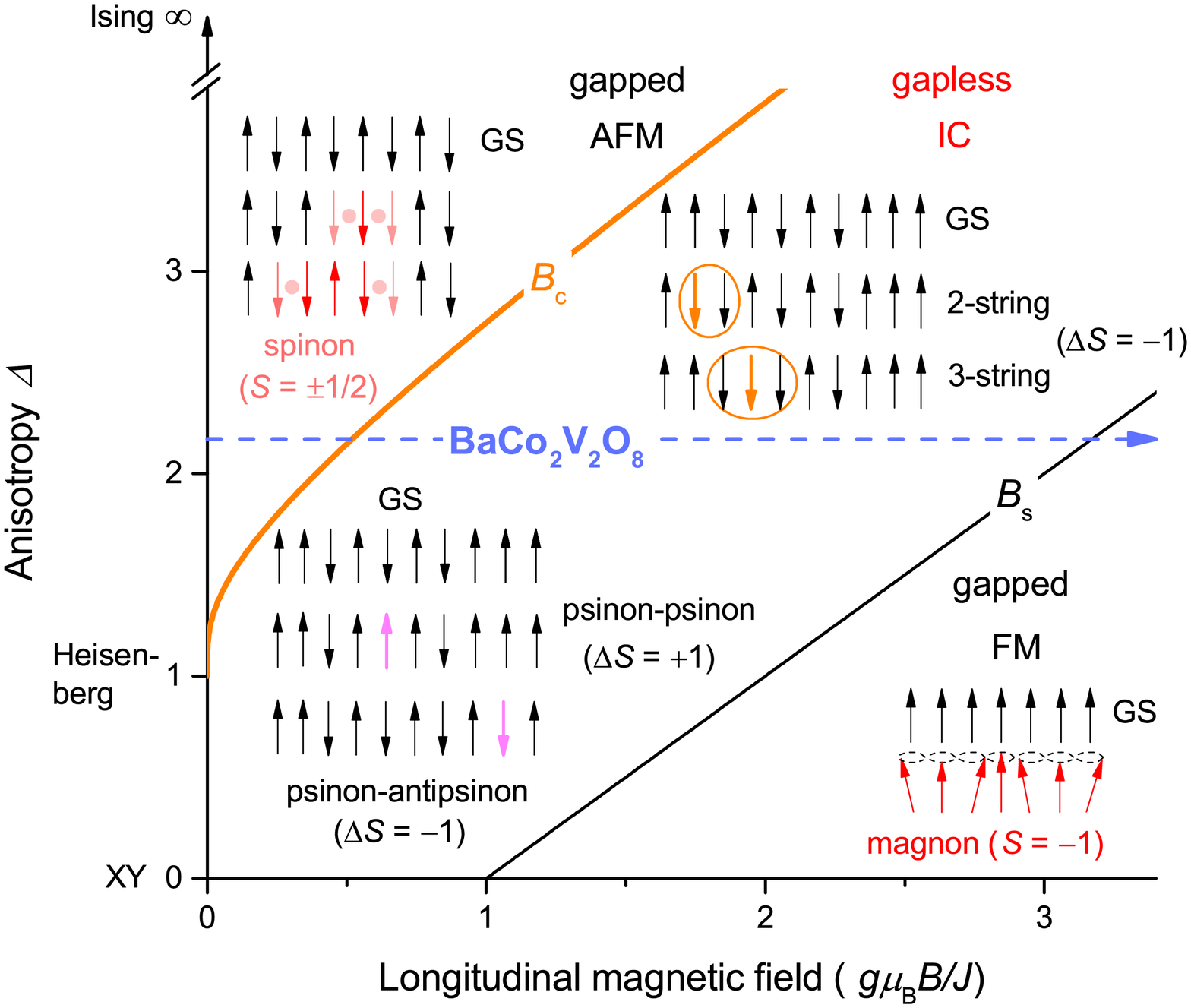}
\vspace{2mm} \caption[]{\label{Fig:phase_diagram}
Phase diagram of the antiferromagnetic XXZ spin-1/2 chain with the anisotropy parameter $\Delta$ in a longitudinal magnetic field.
The different regimes (gapped or gapless) are separated by the critical field $B_c$ and saturation field $B_s$, and are characterized by distinct ground states (GS) and characteristic excitations
(magnons, spinons, psinons, antipsinons, and strings) due to a "spin flip", as schematically illustrated by the insets. The dashed line indicates an increasing magnetic field at $\Delta = 2.17$ for \BaCoVO.
}
\end{figure}

A paradigmatic spin-interaction model to investigate
the quantum critical dynamics is the one-dimensional (1D) spin-1/2 Heisenberg-Ising (or XXZ) model
\begin{equation}\label{Eq:Ising-Heisenberg}
    J\sum_i^N\left[(S^x_i S^x_{i+1}+ S^y_i S^y_{i+1}) + \Delta S^z_i S^z_{i+1}\right]-g\mu_BB\sum_i^N S^z_i,
\end{equation}
with nearest-neighbor antiferromagnetic exchange interaction $J>0$.
In a longitudinal magnetic field, this integrable model allows for
a precise calculation of the many-body spin dynamics \cite{Bethe31,Gaudin71,Takahashi72,Takahashi05,Kitanine99,Karbach00,Caux05,Pereira08,Kohno09,Yang17,Batchelor07} as well as of the ground state \cite{Yang66}.
In the large-Ising-anisotropy ($\Delta>1$) or high-field ($B>B_s$) limit (see Fig.~\ref{Fig:phase_diagram}), the spin dynamics is characterized by gapped excitations on top of an antiferromagnetic or a field-polarized ferromagnetic ground state, respectively.
While the latter has particle-like magnon (or spin wave) excitations of integer quantum number $S = 1$,
the former is featured by a continuum of spinon-pair excitations with fractionalized quantum number $S=1/2$.
As tuning into the gapless regime from different directions,
one can naturally expect richer spin dynamics due to strongly enhanced spin fluctuations, especially for small magnetizations.

Indeed, it has been theoretically shown that in the XY limit $|\Delta|<1$,
not only two-spinon but also four-spinon excitations contribute significantly to the spin dynamics \cite{Caux05}.
For the isotropic Heisenberg model ($\Delta=1$), even high-energy Bethe (or Bethe-Gaudin-Takahashi) string states (complex magnon bound states \cite{Bethe31,Gaudin71,Takahashi72,Takahashi05}) possess certain weight in the transverse sector of the dynamic structure factor \cite{Kohno09}.
Although the multiple fractional excitations were observed by inelastic neutron scattering experiments in various Heisenberg spin chain compounds \cite{Lake05,Stone03,Mourigal13,Lake13},
an observation of the high-energy string excitations remains challenging to a scattering experiment.
Alternatively, by using local quantum quenches, signatures of string states can appear in time-dependent nonequilibrium dynamics \cite{Ganahl12,Liu14}, and were indeed evidenced in a cold-atom experiment for $\Delta\approx1$ \cite{Fukuhara13}.
However, it is far from obvious how to realize the quench dynamics in a condensed-matter experiment.

In the relatively underexplored Ising-like regime ($\Delta>1$), in contrast,
a recent theoretical work \cite{Yang17} found that even in equilibrium, the high-energy string states can have considerable contribution to the spin dynamics in the field-induced gapless phase ($B_c<B<B_s$).
In particular, close to the critical line $B=B_c$, where quantum spin fluctuations are extremely strong,
the string states should be detectable by a scattering experiment.
Our experimental approach here is to carry out high-resolution terahertz (THz) spectroscopic measurements on the Ising-like spin-1/2 chain compound BaCo$_2$V$_2$O$_8$ in an applied longitudinal magnetic field (see Fig.~\ref{Fig:phase_diagram}).
By tracking sharp resonance absorption lines as a function of magnetic field, and comparing it to a Bethe Ansatz calculation,
we unambiguously determine many-body string and fractional spin excitations.
This further allows us to identify the dominant excitations governing the quantum critical dynamics.

Realization of strong spin anisotropy has been found in many Co$^{2+}$ based materials \cite{Lines63,Folen68,Shiba03,He05,Coldea10,Morris14,Wang15a,Wang16,Bera17,Mena18,Wang18a,Breunig13,Breunig15,Faure17}.
Due to spin-orbit coupling and crystal-field effects, interactions between the effective $S=1/2$ Co$^{2+}$ ions in \BaCoVO~can be described
by the 1D antiferromagnetic XXZ model with an Ising-like anisotropy \cite{He05,Kimura07,Niesen13,Niesen14,Klanj15,Grenier15,Wang18}.
High-quality single crystals prepared by the floating-zone
method \cite{Niesen13} were studied by longitudinal-field ($B \parallel c$) magnetization and magnetocaloric-effect measurements, in pulsed fields with a rise time of 7 and 14~ms, respectively, and a pulse duration of 20 and 36~ms.
Quasi-adiabatic conditions were maintained for the magnetocaloric-effect measurements \cite{Kihara13}.

\begin{figure}[b]
\centering
\includegraphics[width=80mm,clip]{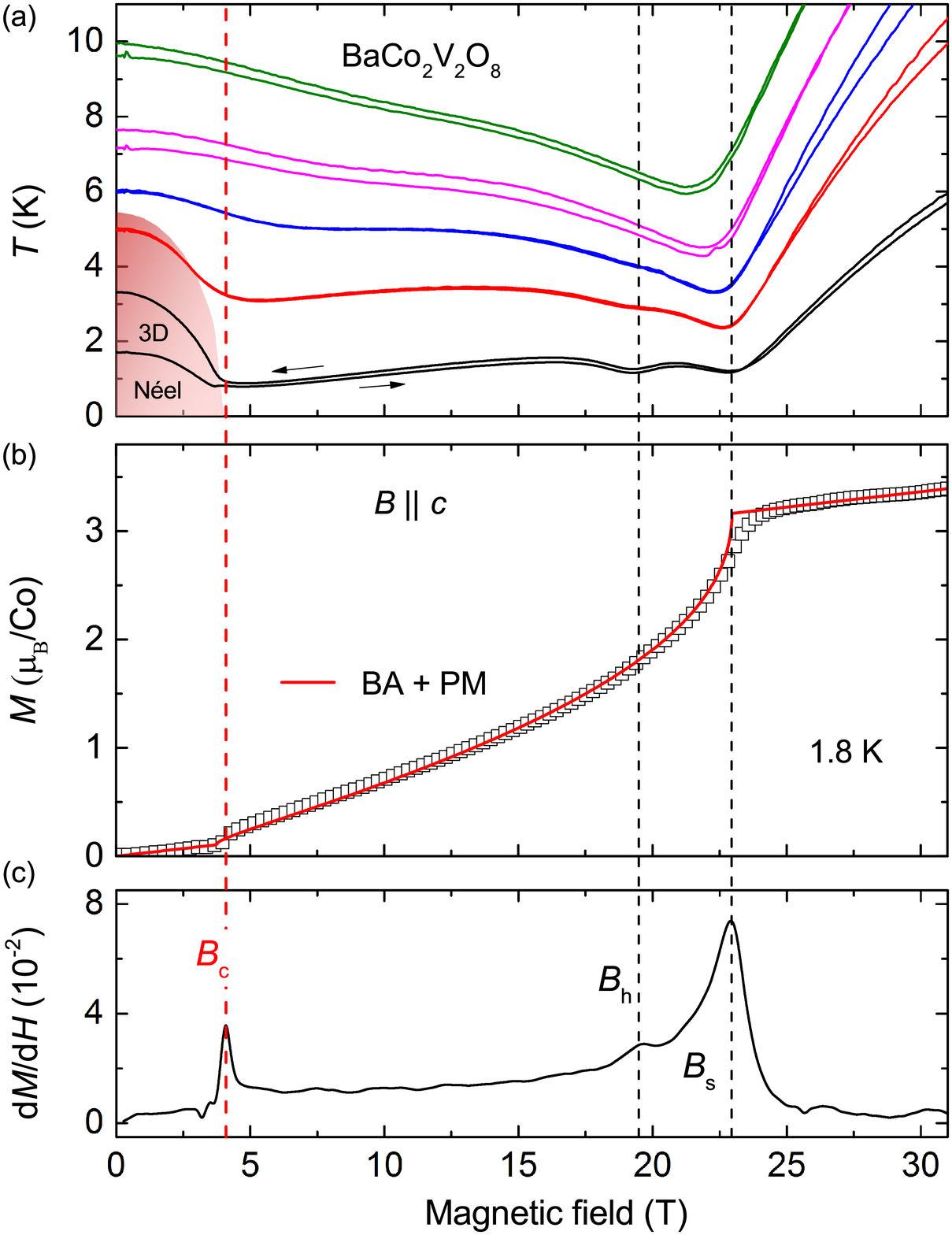}
\vspace{2mm} \caption[]{\label{Fig:MCE_M}
(a) Magnetocaloric-effect $T(B)$, (b) magnetization $M(B)$, and (c) differential susceptibility $dM/dH$ measured as a function of longitudinal field $B \parallel c$.
The critical field $B_c=3.8$~T, and the characteristic fields $B_h=19.5$ and $B_s=22.9$~T corresponding to half-saturated and saturated magnetizations, respectively, yield minima of $T(B)$ and peaks of $dM/dH$ at the lowest temperature, as marked by the dashed lines.
In (a), the shaded area indicates the 3D N\'{e}el ordered phase \cite{Niesen13}.
In (b), the solid line is obtained from Bethe Ansatz (BA) solution of the Heisenberg-Ising model, in addition to a paramagnetic (PM) background.
}
\end{figure}

As shown in Fig.~\ref{Fig:MCE_M}(a), starting from 1.7~K at zero field, the sample temperature $T(B)$ reaches a minimum at $B_c=3.8$~T, which is followed by a slight increase towards higher fields. Before the temperature increases drastically above $B_s=22.9$~T, there is a weak minimum around $B_h=19.5$~T, which corresponds to the half-saturated magnetization [see Fig.~\ref{Fig:MCE_M}(b)] and reflects commensurate fluctuations \cite{Klanj15}.
With down-sweeping the field, $T(B)$ follows very well the up-sweeping curve, apart from a heating effect in the 3D ordered phase.
At elevated temperatures,
the minima at $B_h$ and $B_c$ rapidly broaden and disappear, whereas the minimum at $B_s$ remains sharp and becomes the only dominant feature above $T_N$.
These features qualitatively agree with the phase diagram expected for the 1D Heisenberg-Ising model (Fig.~\ref{Fig:phase_diagram}).
This is also reflected by the magnetization results, which are quantitatively reproduced by the 1D Heisenberg-Ising model with $J = 2.6$~meV, $\Delta=2.17$, and $g = 6.2$, in addition to a weak paramagnetic background with $\chi_{pm}=0.028\mu_B/T$ [solid line, Fig.~\ref{Fig:MCE_M}(b)].
As will be seen below, this parameter set simultaneously provides an excellent description of the observed spin dynamics.

Dynamical properties were obtained by measuring THz transmission spectra on single crystals with typical size $4\times4\times0.5$~mm$^3$,
using a commercial time-resolved spectrometer (Toptica Photonics/TeraFlash).
Dependence on the longitudinal field ($B \parallel c$) was acquired up to 7~T in Faraday geometry with a magneto-optical cryostat (Oxford Instruments/Spectromag).

\begin{figure}[t]
\centering
\includegraphics[width=80mm,clip]{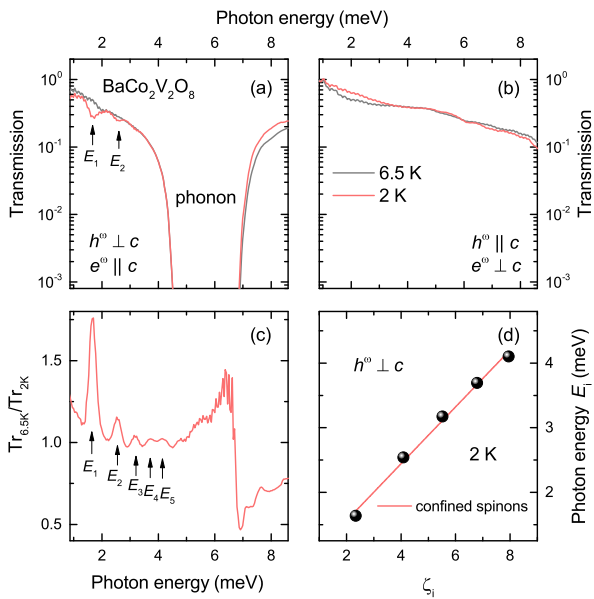}
\vspace{2mm} \caption[]{\label{Fig:Spinon}
Transmission spectra measured at 6.5 and 2~K, above and below $T_N\simeq5.5$~K, respectively, for the polarizations (a) $h^\omega \perp c$, $e^\omega \parallel c$ and (b) $h^\omega \parallel c$, $e^\omega \perp c$.
(c) The ratio of the transmission spectra at 6.5 and 2 K from (a) reveals
absorption peaks, as indicated by the arrows,
which signal the excitations of confined spinons below $T_N$.
(d) The energy hierarchy of these excitations linearly depends on $\zeta_i$, as expected for confined spinons in a linear confining potential \cite{Wang15a}.
}
\end{figure}

Transmission spectra above and below $T_N$ are shown in Fig.~\ref{Fig:Spinon}(a) for $h^\omega \perp c$, $e^\omega \parallel c$, and Fig.~\ref{Fig:Spinon}(b) for $h^\omega \parallel c$, $e^\omega \perp c$.
For $h^\omega \perp c$ and $e^\omega \parallel c$, the spectra are dominated by a strong phonon band between 5 and 7~meV, and at 2~K (below $T_N$) additional absorption peaks are observed at 1.6 and 2.5~meV as indicated by the arrows.
In contrast, in the available spectral range, the spectra for $h^\omega \parallel c$ and $e^\omega \perp c$ are essentially featureless. Therefore, the observed phonon is only active for $e^\omega \parallel c$, while the low-energy excitations are magnetic and active for $h^\omega \perp c$.
The observed selection rule for the magnetic excitations reflects the Ising anisotropy of the antiferromagnetic order \cite{Klanj15,Grenier15}.

\begin{figure}[t]
\centering
\includegraphics[width=70mm,clip]{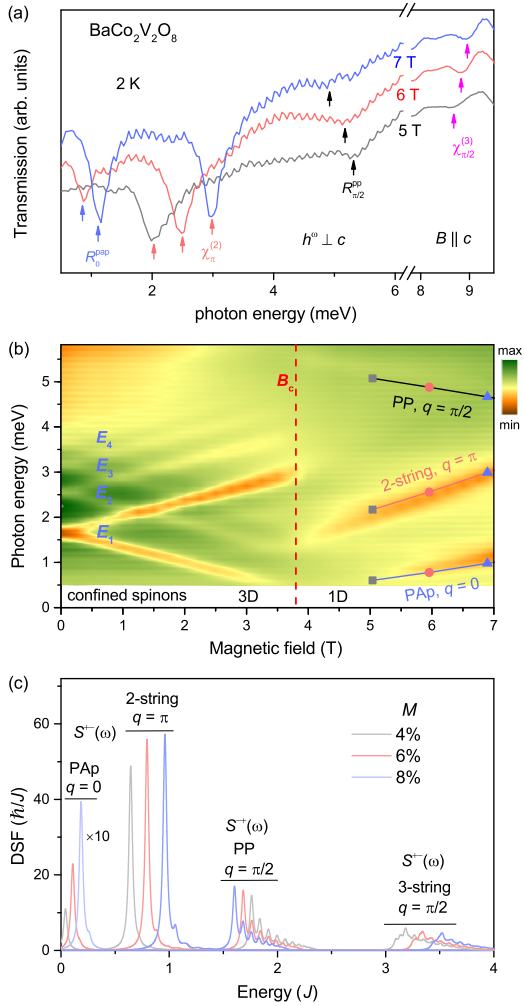}
\vspace{2mm} \caption[]{\label{Fig:String}
(a) Transmission spectra for the polarization $h^\omega \perp c$ obtained at 2~K in the longitudinal fields above $B_c$.
Excitations of 2-strings $\chi_\pi^{(2)}$, 3-strings $\chi_{\pi/2}^{(3)}$, psinon-antipsinon $R_{0}^{pap}$ pairs, and psinon-psinon $R_{\pi/2}^{pp}$ pairs are indicated, as identified by comparing to Bethe Ansatz results.
The weak wiggle-like feature is due to multiple interference within the sample.
(b) Contour plot of transmission at 2~K.
For $B<B_c$, confined-spinon levels $E_i$ are observed.
For $B>B_c$, different excitations with distinct slopes are observed, as for the string and fractional excitations.
(c) Dynamic structure factors (DSF) $S^{+-}$ and $S^{-+}$ obtained via Bethe Ansatz for the magnetizations 2, 4, 6$\%$ of the saturated value.
For clarity, the DSF for PAp is multiplied by a factor of 10.
}
\end{figure}

Besides a broad peak feature at 5 -- 7 meV [Fig.~\ref{Fig:Spinon}(c)], corresponding to a slight temperature-dependent shift of the phonon band,
the ratio of the 6.5 and 2~K transmission spectra exhibits a series of peaks ($E_i$, $i=1,2,3,4,5$) at lower energies, appearing below the magnetic phase transition at $T_N\simeq 5.5$~K.
As expected for the excitations of confined spinons \cite{Kimura07,Coldea10,Morris14,Wang15a,Bera17,Wang16,Mena18}, the energy hierarchy of these modes clearly follows the linear dependence on $\zeta_i$,
$E_i=2E_0+\zeta_i\lambda^{2/3} \left(\hbar^2/\mu\right)^{1/3},  i=1,2,3,...$
where $\zeta_i$ are the negative zeros of the Airy function $Ai(-\zeta_j)=0$, $\mu$ is the effective mass, $\lambda$ measures the confining potential, and $2E_0$ is the threshold energy for creating two spinons \cite{McCoy78}.
The linear fit in Fig.~\ref{Fig:Spinon}(d) gives $2E_0=0.68(10)$~meV and $\lambda^{2/3}\left(\hbar^2/\mu\right)^{1/3}=0.44(2)$~meV.
While the latter is comparable to the value determined at finite momentum transfers by inelastic neutron scattering \cite{Grenier14}, the threshold energy is lower here, because the lower boundary of the spinon-pair continuum has a minimum at the $\Gamma$ point.
Moreover, instead of an overwhelming excitation continuum above $2E_1$,
the observation of the high-lying confined spinons, i.e. $E_4$, $E_5> 2E_1$ [and their field dependence, see Fig.~\ref{Fig:String}(b)] is a distinct fingerprint of confinement effects in an Ising-like spin chain \cite{James18}.

In the longitudinal field, we observe a Zeeman splitting of the confined spinon states, which all follow a similar field dependence towards $B_c$ with a \emph{g}-factor of 5.7(1).
The softening of the lowest-lying confined-spinon level signals the field-induced quantum phase transition to a gapless phase,
as expected for the 1D XXZ model (see Fig.~\ref{Fig:phase_diagram}).
Since these features have been discussed in previous works \cite{Kimura07,Wang15a,Faure17},
we will focus on the spin dynamics above and close to the quantum phase transition at $B_c=3.8$~T.


In the gapless regime above $B_c$, excitations with different field dependencies are revealed,
signaling very different dynamical properties.
As indicated by the arrows in Fig.~\ref{Fig:String}(a), four modes corresponding to transmission minima can be followed at different fields. With increasing field, while the mode $R_{\pi/2}^{pp}$ softens, the frequencies of the modes $R_0^{pap}$, $\chi_{\pi}^{(2)}$, and $\chi_{\pi/2}^{(3)}$ increase.
To compare these experimental results to theory, we performed Bethe-Ansaltz calculations of the transverse dynamic structure factors $S^{a\bar{a}}(q,\omega)=\pi \sum_\mu \left|\langle \mu |S_q^{\bar{a}}| G\rangle \right| ^2 \delta(\omega-E_\mu+E_G)$ \cite{Yang17},
where $S_q^{\pm}=\frac{1}{\sqrt{N}}\sum_N e^{iqn}S_n^{\pm}$ and $\bar{a}=-a$ with $a=+$ or~$-$.
$S_n^{\pm} \equiv S_n^x \pm S_n^z$ is a spin operator acting on the spin site $n$.
The ground state and excited state are expressed by $|G\rangle$ and $|\mu\rangle$ with eigenenergies of $E_G$ and $E_\mu$, respectively.

As the spin chain in \BaCoVO~has a four-step screw symmetry along the chain direction \cite{He05},
we evaluated not only for $q=0$ but also for $q=\pi$ and $\pi/2$ in order to account for the zone-folding effects.
The dominant modes obtained with narrow line shape are shown in Fig.~\ref{Fig:String}(c).
According to the Bethe Ansatz solution,
these are excitations of psinon-antipsinon (PAp) pairs, psinon-psinon (PP) pairs, length-2 strings, and length-3 strings.
As illustrated schematically in the inset of Fig.~\ref{Fig:phase_diagram} ($B_c<B<B_s$),
while the psinons/antipsinons are domain-wall-like fractional excitations,
the 2-string and 3-string excitations are complex magnon bound states of two and three magnons, respectively \cite{Bethe31,Gaudin71,Takahashi72,Takahashi05,Yang17}.
By performing Bethe Ansatz calculations at different magnetizations (or equivalent magnetic fields),
we derive characteristic field dependencies of the various sharp excitations [see Fig.~\ref{Fig:String}(c)],
which allows us to unambiguously identify the experimentally observed excitations.

As shown in Fig.~\ref{Fig:String}(b), the Bethe Ansatz results can be very well compared to the observed excitations above $B_c$, with the same set of parameter as for fitting the magnetization [Fig.~\ref{Fig:MCE_M}(b)].
The resonance frequencies of these modes follow very nicely the theoretically expected linear dependence on the magnetic field above 5~T. To take into account the nonlinear dependence approaching $B_c$, the theory results are shifted upward by a constant of $0.5$~meV, which is smaller than the lower energy limit of our spectroscopy.
The overall excellent comparison allows us to identify the experimentally observed modes as fractional psinon/antipsinon excitations as well as 2-string and 3-string excitations, as marked in Fig.~\ref{Fig:String}(a)(b). The revealed selection rule is summarized in Table~\ref{tab:sel_rule}.

\begin{table}[h]
\caption{\label{tab:sel_rule}  Selection rule for the observed confined spinons, psinon-psinon (PP), psinon-antipsinon (PAp), and string excitations.}
\begin{ruledtabular}\vspace{0.5mm}
\begin{tabular}{c@{\hspace{2em}}*{5}{c}}
&$H \parallel c$  &$\Delta S$ &$h^\omega \perp c$ &$h^\omega \parallel c$
\\\hline
&confined spinons  &$\pm 1$ &yes  &no
\\
&PP    &$+ 1$    &yes  &n/a
\\
&PAp &$- 1$   &yes  &n/a
\\
&2-string &$- 1$  &yes  &n/a
\\
&3-string &$- 1$  &yes  &n/a
\end{tabular}
\end{ruledtabular}
\end{table}

Comparing the intensities with the fractional excitations, the 2-string mode turns out to have unique features.
From Fig.~\ref{Fig:String}(a), it is, first of all, obvious that the PA and 2-string modes are much sharper and stronger than the PP and 3-string modes.
Secondly, with decreasing field, the intensity of the low-energy PAp modes significantly decreases,
while the field dependence of the 2-string mode is less pronounced.
As a consequence, approaching $B_c$ from above, the 2-string mode becomes finally dominant over the low-energy PA mode.
These features are in good agreement with the theoretical findings [see Fig.~\ref{Fig:String}(c)] \cite{Yang17} as well as with the experimental observations in a related compound \SrCoVO~with weaker Ising anisotropy \cite{Wang18a}. The finding of dominant high-energy string excitations on approaching $B_c$ is in contrast to the conventional understanding that low-energy excitations are dynamically most important.



To conclude, we have identified exotic spin excitations, such as many-body string and fractional excitations, and revealed their selection rule in the vicinity of a field-induced quantum phase transition in the Ising-like spin-1/2 chain compound BaCo$_2$V$_2$O$_8$. Below the phase transition the spin dynamics is characterized by gapped fractional spinon excitations with strong confinement features. In contrast, above the phase transition in the gapless phase, the dominant role is played by the high-energy string excitations although low-energy fractional excitations of psinons/antispinons are present as well. Our results provide a guide for hunting high-energy string excitations, and suggest that the high-energy many-body excitations may generally play an important role in quantum critical dynamics.


\begin{acknowledgements}
We would like to thank Congjun Wu for fruitful discussions.
We acknowledge partial support by the
DFG (German Research Foundation)
via TRR 80: From Electronic Correlations to Functionality (Subproject F5),
and via the project No. 277146847 - CRC 1238: Control and Dynamics of Quantum Materials (Subprojects A02 and B01).
The high field experiments at Dresden were supported by HLD at HZDR, member of the European Magnetic Field Laboratory (EMFL).
Z.W. acknowledges hospitality of ISSP, the University of Tokyo, and of Tsung-Dao Lee Institute.
\end{acknowledgements}

\end{document}